# Constraints on the Power Spectrum of Mass Fluctuations from Galaxy Cluster Peculiar Velocities.

Rupert Croft and George Efstathiou,
*Department of Physics, University of Oxford, Keble Road, Oxford, OX1 3RH.*

## ABSTRACT

We investigate the feasibility of carrying out likelihood analysis on the velocities of galaxy clusters to determine power spectrum parameters. Using N-body simulations of cosmological density fields we show that the velocity field traced by clusters is highly non-linear on small scales ($r < 10$ h$^{-1}$Mpc), mostly due to strong infall into superclusters. The transverse and line of sight velocity correlation functions relevant to likelihood analysis deviate significantly from the linear theory values. Unfortunately, the sparseness of cluster samples means that they are not amenable to any rigorous smoothing procedure to rectify this. We are forced instead to remove close pairs of clusters which would contribute disproportionately to the likelihood. This reduces the size still further of the small samples of observed clusters available to us, forcing us into the conclusion that using straightfoward linear theory covariance analysis on clusters is unpromising. Nevertheless we apply the technique to two sets of observational data and find marginal evidence for more power in the mass density field than standard CDM. The normalisation is consistent with $\Omega^{0.6}\sigma_8 \sim 0.7$ for both data sets, within the (large) errors.

## 1. Introduction

In linear theory, two-point statistics can be used to define the properties of the velocity field entirely. If it were possible to find tracers of cosmological velocity fields that obey this criterion, then covariance analysis could be used to remove uncertainties due to sample geometry[1,2]. Clusters are potentially good tracers because velocities of galaxies in clusters can be averaged to give smaller errors.

## 2. Comparing simulations of cluster formation with linear theory.

### 2.1. Catalogues of clusters from N-Body models.

A series of simulations were carried out in order to assess how well clusters trace the linear velocity field. A P$^3$M N-body code[3] was used to follow the evolution of cold dark matter and clusters were selected from the simulations by a percolation technique. This is as descibed in [4], except that here we have simulated a box size of size 600 h$^{-1}$Mpc (where H$_0$=100h kms$^{-1}$Mpc$^{-1}$) with $4 \times 10^6$ particles. The initial power spectra (hereafter PS) are those of [5], with the shape parametrised by $\Gamma$, equal to $\Omega_0 h$ in CDM models. We will present results based on 2 simulations of the Standard CDM scenario ( for SCDM, $\Omega_0 = 1, \Gamma = 0.5$) and 4 simulations of Low Density CDM (LCDM has $\Omega_0 = 0.2$, $\Gamma = 0.2$, and a cosmological constant $\Lambda$ such that $\Lambda/3H_0^2 = 1 - \Omega_0$). The models are normalised so that the rms mass fluctuations in 8 h$^{-1}$ Mpc spheres, $\sigma_8 = 1.0$ at the present day. Because we choose to study velocities of rich systems, a lower mass bound is applied so that the clusters have a mean separation of 30 h$^{-1}$Mpc.

### 2.2. Velocity correlations.

Although the rms velocities of simulated clusters agree roughly with linear theory[4], approximately 5% of clusters occupy a significant non-Maxwellian tail. Examination of the position of these objects in space shows many of them to be infalling into superclusters. To find the range of scales over which these non-linear effects may be important we examine the velocity correlation tensor, $\Psi_{ij}(\mathbf{r})$. This describes the covariance of different velocity components as a function of distance between two points. It can be written as a sum of the radial ($\Psi_{\parallel}(r)$) and transverse ($\Psi_{\perp}(r)$) velocity correlation functions[6]:

$$\Psi_{ij}(\mathbf{r}) = \langle V_i(\mathbf{x})V_j(\mathbf{x}+\mathbf{r})\rangle = \Psi_{\perp}(r)\delta_{ij} + [\Psi_{\parallel}(r) - \Psi_{\perp}(r)]\hat{r}_i\hat{r}_j. \tag{1}$$

In linear pertubation theory these velocity correlations are given by integrals over the PS[6]:

$$\Psi_{\parallel}(r) = \frac{H_0^2 \Omega^{1.2}}{2\pi^2} \int_0^{\infty} \frac{P(k)}{k^2} \exp(-k^2 R_f^2)(j_0(kr) - \frac{2j_0(kr)}{kr})dk, \tag{2}$$

$$\Psi_{\perp}(r) = \frac{H_0^2 \Omega^{1.2}}{2\pi^2} \int_0^{\infty} \frac{P(k)}{k^2} \exp(-k^2 R_f^2)(\frac{j_1(kr)}{kr})dk, \tag{3}$$

where $j_n$ are the usual spherical Bessel functions, and the velocity field has been filtered with a Gaussian window of radius $R_f$. We will use $R_f = 3$h$^{-1}$mpc in our comparison of linear theory with the velocity correlations of simulated clusters, as this value provided the best fit to the one-dimensional velocity distribution[4].

We have evaluated $\Psi_{\parallel}(r)$ and $\Psi_{\perp}(r)$ for samples of simulated clusters, and the results are shown in Fig. 1, together with the linear theory predictions. It can be seen that cluster pairs with separations $< 10h^{-1}Mpc$ show a strong tendency to fall toward each other ($\Psi_{\parallel}(r)$ is stongly negative) rather than streaming together as expected in linear theory.

### 2.3. Maximum likelihood analysis.

The relative probablility of observing a set of $N$ line-of-sight peculiar velocities $V_i$ at positions $\mathbf{r_i}$ in a Gaussian field is given by a multivariate Gaussian distribution[1]:

$$P(V_i)d^N V = \frac{d^N V}{(2\pi)^{N/2} |C|^{1/2}} \exp(-\frac{1}{2} V_i C_{ij}^{-1} V_j), \tag{4}$$

where $C_{ij} = \langle V_i V j \rangle$ is the covariance matrix of the sample. In linear theory, the terms in $C_{ij}$ are simple functions of the velocity correlations defined in equations 2 and 3. If clusters 1 and 2 are positioned at $\mathbf{r_1}$ and $\mathbf{r_2}$ relative to the observer, and the angles $\theta_1$ and $\theta_2$ are defined by $\cos\theta_1 = \hat{\mathbf{r}}_1 \cdot \hat{\mathbf{r}}_{12}$ and $\cos\theta_2 = \hat{\mathbf{r}}_2 \cdot \hat{\mathbf{r}}_{12}$ then

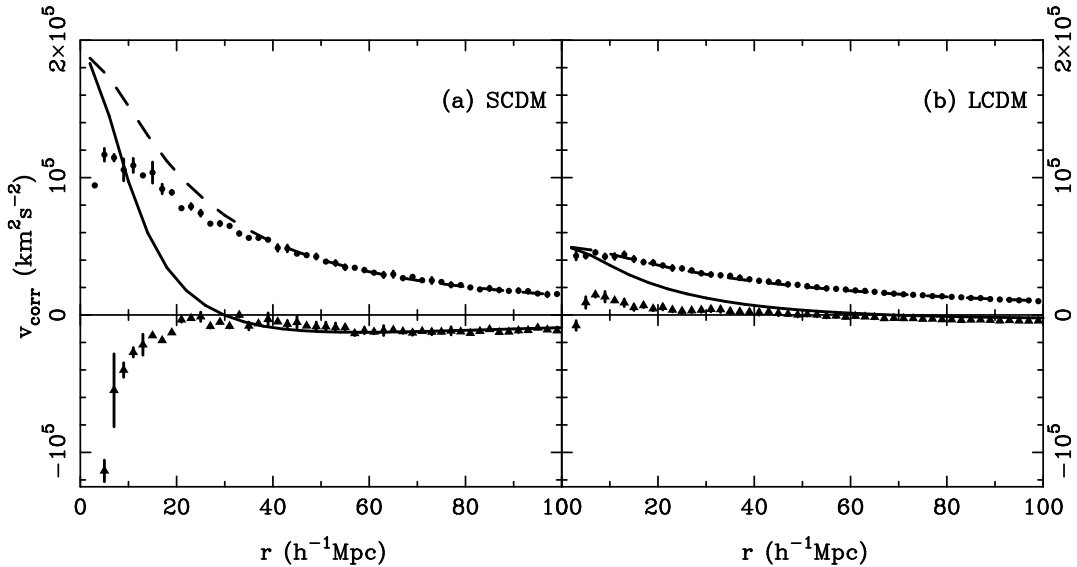

Figure 1: The transverse (upper lines) and parallel (lower lines) velocity correlation functions for rich clusters in (a) the Standard CDM model and (b) Low Density CDM. The smooth curves are the linear theory prediction (equations 2 and 3) and the points are evaluated from samples of simulated clusters.

$$C_{12} = \cos\theta_1 cos\theta_2 \Psi_\|(r_{12}) + \sin\theta_1 sin\theta_2 \Psi_\perp(r_{12}). \qquad (5)$$

For a real sample with velocity errors, there will also be additional contributions to the covariance matrix. We will assume that these errors are uncorrelated and have a normal distribution, so that only the $C_{ii}$ terms are affected, each being increased by $\sigma_i^2$, the estimated error on the velocity of cluster $i$.

The PS used in the calculation of $\Psi_\|$ and $\Psi_\perp(r)$ is again that of [5]. We maximise the likelihood given by equation 4 in the parameter space of $\Gamma$ and effective amplitude $\Omega^{0.6}\sigma_8$. Whilst it would be desirable to test a fit to the PS which is a sum of step functions, this way of finding a more general answer is not practicable with this method, which involves many inversions of the matrix $C$. We are also limited by the small number of data points. The parameter $\Gamma$ should be a fair phenomenological fit to the PS of most reasonable models, though, and we would expect its value to lie somewhere around 0.2, that inferred from the shape of the galaxy PS[7,8].

Testing the procedure on samples of linear theory velocity fields reveals that although there is rather more scatter in the shape parameter than amplitude, the method recovers the right values on average. We then carry out likelihood analysis on samples of simulated clusters. Disappointingly, we find the method unable recover the correct parameters. Tests have been carried out[9] which show that clusters in close pairs dominate the likelihood and skew the parameter determinations to wildly incorrect values. Some means of mitigating the non-linear effects seen on small scales needs to be applied. Unfortunately we find that clumping the data can introduce large biases[9]. By considering grouped data as single objects we are removing small scale power from the velocity field. Unless we can smooth rigorously with a filter that can be introduced into the theoretical covariances of equation 5, we will recover values of $\Gamma$ and $\sigma_8$ which are systematically and unpredictably low. We therefore choose to excise clusters with close neighbours ($r < 10h^{-1}Mpc$) from the calculation altogether. As a final crude attempt to remove non-linearities, we remove the fastest 3% of clusters on the basis of their position in the one-dimensional velocity distribution. The distribution of best fit PS parameters for tests under these conditions is shown in Fig. 2. It is evident that some biases are still present which should be borne in mind later. We have also tried adding simulated observational errors. The results (which are not shown) show that scatter is increased but do not bias the results, as long as we use the real space separations of clusters.

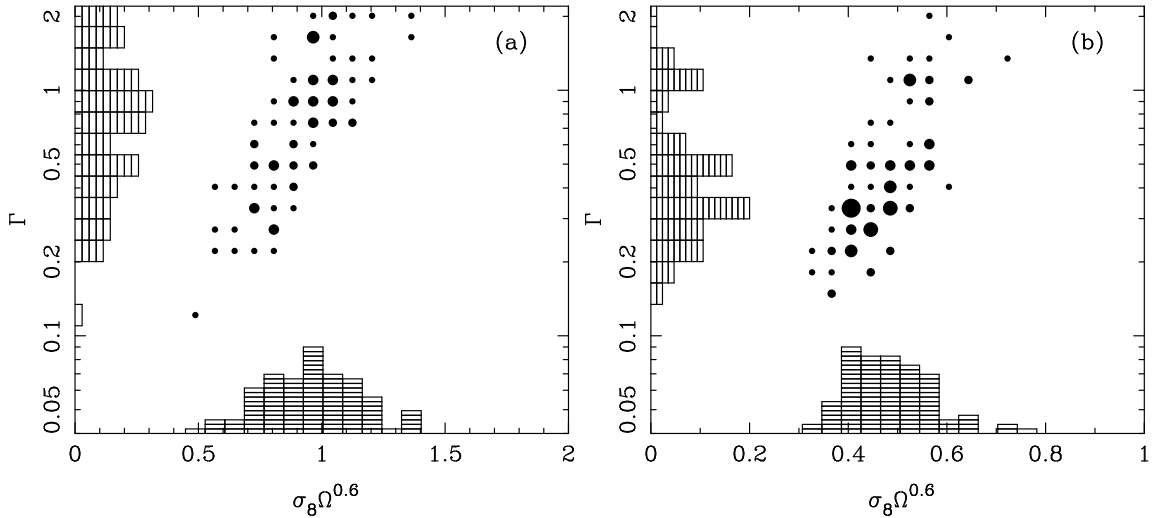

Figure 2: The distribution of best fit values in $\Gamma - \sigma_8$ space found by likelihood analysis of 200 samples of simulated cluster velocity fields (60 clusters each in 100 h$^{-1}$Mpc spheres). Panel (a) is for SCDM ($\Gamma = 0.5, \Omega^{0.6}\sigma_8 = 1$) and (b) is LCDM ($\Gamma = 0.2, \Omega^{0.6}\sigma_8 = 0.38$). Clusters with a nearest neighbour within 10 h$^{-1}$Mpc have been removed from the sample, as well as the fastest 3 per cent of clusters.

## 3. Applying maxmimum likelihood analysis to observed cluster velocities.

*3.1. The cluster samples.*

We deal here with two separate samples of observed clusters, one of which (AFFHM) is a compilation of 65 clusters[10,11,12,13,14] and is described in [4]. The other (LP) consists of distance and velocity measurements of all 119 Abell clusters within 150 h$^{-1}$mpc using the Brightest Cluster Galaxy method[15]. These distances have fairly large (16%) errors, as only one measurement is possible per cluster. On the other hand, the large volume occupied by the sample has already been used to infer a bulk flow which has significant power to rule out models[16,17,2].

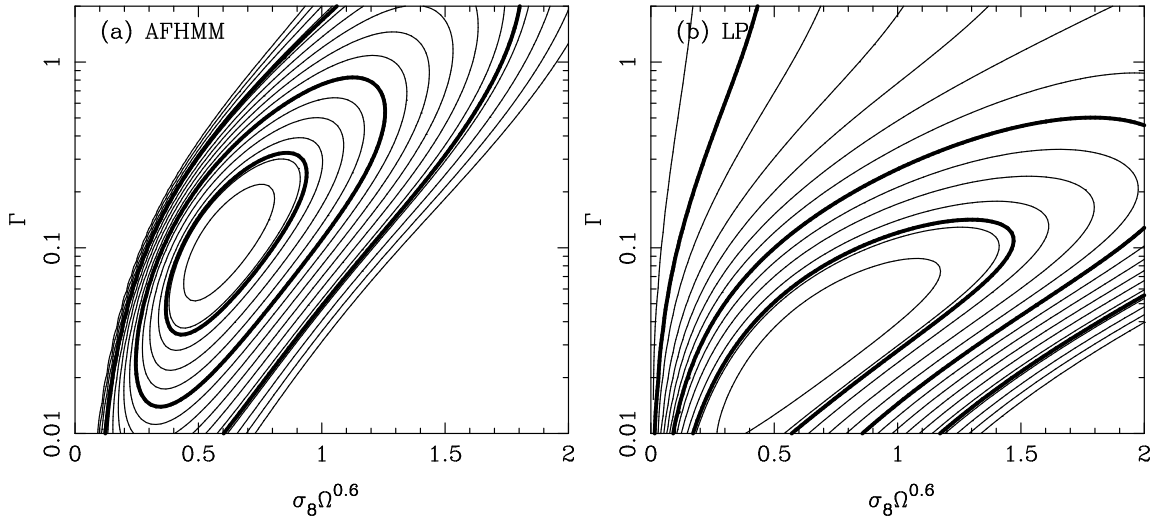

Figure 3: Contour plots resulting from likelihood analysis of the cluster velocities in (a) the combined AFHMM sample and (b) the Lauer & Postman sample. The heavy contours enclose 68 per cent, 95 per cent and 99.7 per cent of the probablity if the distribution of $-2\ln(\text{likelihood})$ follows a $\chi^2$ distribution.

*3.2. Results.*

Applying the same treatment used on the simulated clusters in Sec. 2 to these data sets, we are left with only 38 and 89 objects in the AFFHM and LP sets respectively. We then calculate the likelihoods resulting from equation 0.6 on a dense grid of $\Gamma$ and $\Omega^{0.6}\sigma_8$ values. The results are shown in Fig. 3 with the heavy lines representing confidence limits that the result lies within those contours. Both samples point to a best fit amplitude of $\Omega^{0.6}\sigma_8 \sim 0.7$ and to a PS with more large scale power than SCDM ($\Gamma \sim 0.1$ for AFHHM and $\sim 0.05$ for LP). If we assume that the PS shape is known to be 0.2 as derived from measurements of the galaxy distribution, then maximising the likelihood for variations in only $\Omega^{0.6}\sigma_8$ gives $\Omega^{0.6}\sigma_8 = 0.73^{+0.33}_{-0.27}$ for AFHHM and $\Omega^{0.6}\sigma_8 = 1.44^{+0.74}_{-0.65}$ for L&P (90% confidence limits).

## 4. Conclusions

We find that simulated clusters show strong non-linear infall into high density regions. Clusters are distributed too sparsely to allow for smoothing with a well defined filter in an attempt to reduce the non-linear effects. Any crude grouping of the data will introduce strong biases into the determination of power spectrum parameters. Rather than abandon linear theory covariance analysis altogether, we have chosen to remove high peculiar velocity clusters from the likelihood calculation. Applying the technique to two different samples of observed clusters results in determinations of $\Gamma$ and $\sigma_8$ which are consistent ($\Gamma \sim 0.05 - 0.3$ and $\sigma_8 \sim 0.4 - 1.0$). Taking into consideration possible remaining systematic errors, both the SCDM and LCDM models appear to lie $\sim 1$ sigma away from the best fit power spectrum, which itself is near to that of the Mixed Dark Matter model[18].

## 5. Acknowledgements

RACC acknowledges the receipt of a SERC/PPARC studentship and thanks the conference organisers for financial support.